\documentclass[pra,aps,showpacs,superscriptaddress,twocolumn]{revtex4}

\usepackage{amsmath}
\usepackage{bm}
\usepackage{graphicx}

\begin{document}

\title{Penetration of a vortex dipole across an interface of Bose--Einstein
condensates}

\author{Tomohiko Aioi}
\affiliation{Department of Engineering Science, University of
Electro-Communications, Tokyo 182-8585, Japan}

\author{Tsuyoshi Kadokura}
\affiliation{Department of Engineering Science, University of
Electro-Communications, Tokyo 182-8585, Japan}

\author{Hiroki Saito}
\affiliation{Department of Engineering Science, University of
Electro-Communications, Tokyo 182-8585, Japan}

\date{\today}

\begin{abstract}
The dynamics of a vortex dipole in a quasi-two dimensional two-component
Bose--Einstein condensate are investigated.
A vortex dipole is shown to penetrate the interface between the two
components when the incident velocity is sufficiently large.
A vortex dipole can also disappear or disintegrate at the interface
depending on its velocity and the interaction parameters.
\end{abstract}

\pacs{03.75.Mn, 03.75.Lm, 67.85.De, 67.85.Fg}

\maketitle

\section{Introduction}

Vortex rings, such as smoke rings blown from smoker's mouth, propagate in
the direction of the axis of the ring with a long lifetime.
The two-dimensional (2D) analogue of a vortex ring is a vortex dipole,
which consists of a pair of vortices of opposite circulations, 
propagating side by side.
A vortex dipole is a very stable structure and carries momentum in the
direction of propagation.

In superfluids, a vortex dipole consists of a pair of quantized vortices,
whose circulations are quantized to $\pm h / m$ with $h$ being Planck's
constant and $m$ being an atomic mass.
Such quantized vortex dipoles have been generated and observed in
Bose--Einstein condensates (BECs) of atomic
gases~\cite{Inouye,Neely,Freilich,Middel} and of exciton
polaritons~\cite{Roumpos,Nardin}.
A quantized vortex dipole propagates with velocity $\simeq \hbar / (m d)$,
where $d$ is the distance between vortices.
In a uniform superfluid, $d$ is constant in time and a quantized vortex
dipole propagates with a constant velocity.
In an inhomogeneous system, on the other hand, each vortex of a vortex
dipole follows its own trajectory~\cite{Neely,Freilich,Middel} or even
remains stationary~\cite{Freilich,Crasovan}.

In the experiment reported in Ref.~\cite{Neely}, a vortex dipole was
generated in an oblate BEC, which propagated through the system.
When a vortex dipole reached the edge of the condensate, the vortex and
antivortex disintegrated and they moved along the edge of the condensate
in opposite directions.
This behavior is similar to that of a classical vortex dipole moving
towards a rigid wall~\cite{Lamb}, where the vortex and antivortex
disintegrate and move along the wall in opposite directions.
For both a wall and the edge of a condensate, a vortex dipole cannot go
beyond the boundary.
In this paper, we investigate the dynamics of a vortex dipole moving
towards an interface between BECs of different components.
Few studies have been performed on this kind of vortex dynamics at
interfaces even in classical fluids.
In classical fluids, dynamics of vortex rings moving towards a density
interface have been studied~\cite{Linden,Dahm,Kuehn}.

In the present paper, we will show that a vortex dipole can penetrate the
interface in a two-component BEC, in which quantized vortices in one
component are transferred to the other component.
A vortex dipole also disappears or disintegrates at the interface.
These behaviors depend on the incident velocity of the vortex dipole and
the atomic scattering lengths that determine the interfacial tension and
the width of the interface.
For some parameters, the cores of the vortex dipole after the penetration
are occupied by the other component.
When only one vortex is occupied, the occupying fraction tunnels and
oscillates between the vortex pair.
In a trapping potential, a rich variety of vortex dynamics can be
observed.

This paper is organized as follows.
Section~\ref{s:form} gives a basic formalism.
Section~\ref{s:ideal} numerically demonstrates the dynamics of vortex
dipoles in an ideal system and discusses the parameter dependence of the
dynamics.
Section~\ref{s:trap} investigates the dynamics of vortex dipoles in a
trapped BEC.
Section~\ref{s:conc} provides conclusions to this study.

\section{Formulation of the problem}
\label{s:form}

We consider a two-component BEC in mean-field theory.
The dynamics of the macroscopic wave functions $\psi_1$ and $\psi_2$ of
components 1 and 2 are described by the Gross--Pitaevskii (GP) equation,
\begin{subequations} \label{GP}
\begin{eqnarray}
i \hbar \frac{\partial{\Psi_1}}{\partial t} & = & \left( -\frac{\hbar^2}{2
m_1} \nabla^2 + V_1 + G_{11} |\Psi_1|^2 + G_{12} |\Psi_2|^2 \right)
\Psi_1, \nonumber \\
\\
i \hbar \frac{\partial{\Psi_2}}{\partial t} & = & \left( -\frac{\hbar^2}{2
m_2} \nabla^2 + V_2 + G_{22} |\Psi_2|^2 + G_{12} |\Psi_1|^2 \right)
\Psi_2, \nonumber \\
\end{eqnarray}
\end{subequations}
where $m_j$ is an atomic mass, $V_j$ is an external potential, and
$G_{jj'} = 2 \pi \hbar^2 a_{jj'} (m_j^{-1} + m_{j'}^{-1})$ with $a_{jj'}$
being the $s$-wave scattering lengths between atoms in components $j$ and
$j'$ ($j, j' = 1, 2$).
The wave function is normalized as $\int |\Psi_j|^2 d\bm{r} = N_j$, where
$N_j$ is the number of atoms in component $j$.

For simplicity, we restrict ourselves to quasi-2D systems in the following
analysis.
The system is assumed to be tightly confined in the $z$ direction by a
potential $V_{zj}(z)$ and the wave function is reduced to the form
$\Psi_j(\bm{r}) = \phi_j(z) \psi_j(x, y)$, where $\phi_j(z)$ is the
normalized wave function of the ground state for the potential
$V_{zj}(z)$.
Multiplying $\phi_j(z)$ to Eq.~(\ref{GP}) and integrating it with respect
to z, we have the effective 2D GP equation,
\begin{subequations} \label{2DGP}
\begin{eqnarray}
i \hbar \frac{\partial{\psi_1}}{\partial t} & = & \left( -\frac{\hbar^2}{2
m_1} \nabla_\perp^2 + V_{\perp 1} + g_{11} |\psi_1|^2 + g_{12} |\psi_2|^2
\right) \psi_1, \nonumber \\
\\
i \hbar \frac{\partial{\psi_2}}{\partial t} & = & \left( -\frac{\hbar^2}{2
m_2} \nabla_\perp^2 + V_{\perp 2} + g_{22} |\psi_2|^2 + g_{12} |\psi_1|^2
\right) \psi_2,
\nonumber \\
\end{eqnarray}
\end{subequations}
where $\nabla_\perp^2$ is the 2D Laplacian, $V_{\perp j}(x, y) =
V_j(\bm{r}) - V_{zj}(z)$, and 
\begin{equation} \label{geff}
g_{jj'} = G_{jj'} \int |\phi_j|^2 |\phi_{j'}|^2 dz.
\end{equation}
We assume that the interaction parameters $g_{jj'}$ satisfy the immiscible
condition,
\begin{equation}
g_{11} g_{22} < g_{12}^2.
\end{equation}

We solve the effective 2D GP equation (\ref{2DGP}) numerically.
The initial state is the ground state obtained by the imaginary-time
propagation of Eq.~(\ref{2DGP}), i.e., $i$ in the left-hand sides of
Eq.~(\ref{2DGP}) is replaced with $-1$.
The imaginary- and real-time propagations are obtained using the
pseudospectral method~\cite{Recipes}.
The computational size is large enough that the boundary condition does
not affect the results.

\section{Dynamics of vortex dipoles in an ideal system}
\label{s:ideal}

We first consider an ideal system to study the dynamics of vortex dipoles,
where the trapping potential is absent and the interface between the two
components is straight along the $y$ axis.
For simplicity, we assume $m_1 = m_2$ and $g_{11} = g_{22} \equiv g$.
The width of the interface is characterized by the parameter
\begin{equation}
\Delta \equiv \frac{g_{12}}{g} - 1.
\end{equation}
For $0 < \Delta \ll 1$, the density distribution with the boundary
condition $\lim_{x \rightarrow -\infty} n_1 = \lim_{x \rightarrow \infty}
n_2 = n_0$ is given by~\cite{Barankov}
\begin{equation}
n_j(x) \simeq \frac{n_0}{2} \left[ 1 + (-1)^j \tanh \frac{\sqrt{2 \Delta}
x}{\xi} \right],
\end{equation}
where $n_0$ is the density far from the interface and $\xi = \hbar / (m g
n_0)^{1/2}$ is the healing length.
The width of the interface $w$ is thus proportional to $\xi /
\sqrt{\Delta}$.
The time is normalized as $\tilde t = t v_{\rm s} / \xi$, where $v_{\rm
s} = (g n_0 / m)^{1/2}$ is the sound velocity.

A vortex dipole is generated by the method proposed in Ref.~\cite{Aioi}
(see Fig.~3 in Ref.~\cite{Aioi}).
When an attractive Gaussian potential produced by a red-detuned laser beam
is displaced in a quasi-2D BEC, a vortex dipole is created in front of the
potential and ``launched'' in the direction of the
displacement~\cite{Aioi}.
We create a vortex dipole in component 1 using a Gaussian potential with
magnitude $V_0 > 0$ and width $w$ as
\begin{equation} \label{Gaussian}
V_{\perp 1} = V_{\rm G} \equiv -V_0 \exp \left\{ -\frac{[\bm{r} -
\bm{r}_0(t)]^2}{w^2} \right\}.
\end{equation}
The position $\bm{r}_0(t)$ is linearly displaced between the times $t = 0$
and $t = T$, and a vortex dipole is launched from the potential towards
the direction of $\bm{r}_0(T) - \bm{r}_0(0)$.
The velocity of a vortex dipole can be controlled by the parameters $V_0$,
$w$, and the function $\bm{r}_0(t)$.
The Gaussian potential is located far from the interface, and its motion
does not affect the dynamics near the interface.
The dynamics of a vortex dipole therefore depend only on its velocity and
not on each parameter in Eq.~(\ref{Gaussian}).

\begin{figure}[t]
\includegraphics[width=8cm]{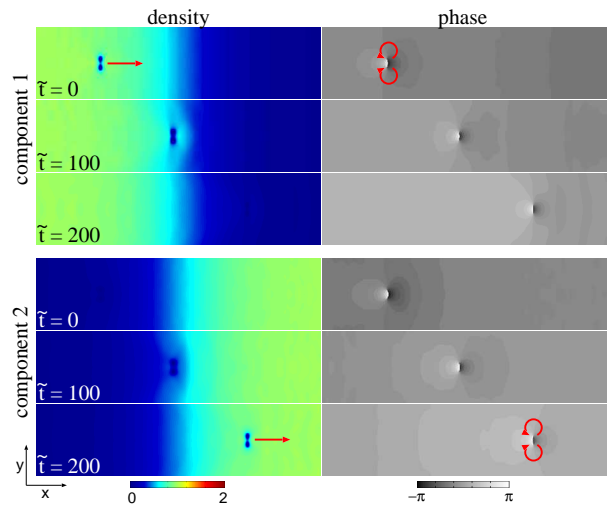}
\caption{
(Color online) Dynamics of normalized density profiles $|\psi_j|^2 / n_0$
and phase profiles ${\rm arg}\, \psi_j$ for $g_{11} = g_{22}$ and $\Delta =
g_{12} / g_{11} - 1 = 10^{-3}$.
The velocity of the incident vortex dipole is $v_{\rm in} = 0.31 v_{\rm
s}$, where $v_{\rm s}$ is the sound velocity. 
The direction of the vortex dipole propagation is indicated by the arrows
in the left panels, and the circulations of vortices are indicated by the
arrows in the right panels.
The time is normalized as $\tilde t = t v_{\rm s} / \xi$, where $\xi$ is
the healing length.
The origin of time $\tilde t = 0$ is taken arbitrarily.
The field of view of each panel is $120 \xi \times 30 \xi$.
}
\label{f:penet1}
\end{figure}
Figure~\ref{f:penet1} demonstrates the typical penetration dynamics of a
vortex dipole through an interface.
A vortex dipole is generated in component 1 and propagates in the $+x$
direction, which is perpendicular to the interface.
We note that the vortex dipole in component 1 is accompanied by a ``ghost
vortex dipole'' in component 2, and they are located in the same position
(${\rm arg}\, \psi_1$ and ${\rm arg}\, \psi_2$ at $\tilde t = 0$ in
Fig.~\ref{f:penet1}).
As the vortex dipole propagates through the interface region, the ghost
vortex dipole in component 2 is substantiated ($|\psi_2|^2 / n_0$ at
$\tilde t = 100$ in Fig.~\ref{f:penet1}) and then the vortex dipole is
completely transferred to component 2 ($|\psi_2|^2 / n_0$ at $\tilde t =
200$ in Fig.~\ref{f:penet1}).
After passing through the interface, the vortex dipole in component 1
becomes a ghost (${\rm arg}\, \psi_1$ at $\tilde t = 200$ in
Fig.~\ref{f:penet1}).

\begin{figure}[t]
\includegraphics[width=8cm]{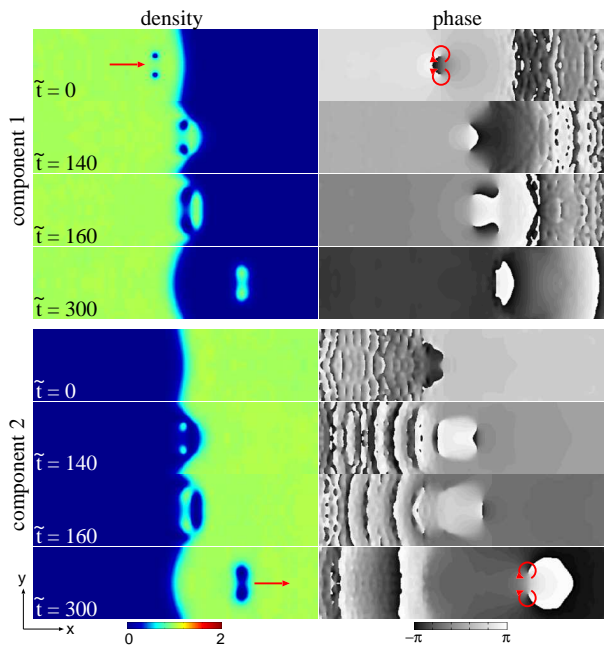}
\caption{
(Color online) Dynamics of normalized density profiles $|\psi_j|^2 / n_0$
and phase profiles ${\rm arg} \, \psi_j$ for $\Delta = 0.05$ and $v_{\rm
in} = 0.12 v_{\rm s}$.
The direction of the vortex dipole propagation is indicated by the arrows
in the left panels, and the circulations of vortices are indicated by the
arrows in the right panels.
The field of view of each panel is $120 \xi \times 30 \xi$.
}
\label{f:penet2}
\end{figure}
Figure~\ref{f:penet2} shows the case of a thinner interface (larger
$\Delta$) and a slower vortex dipole compared with those in
Fig.~\ref{f:penet1}.
When a vortex dipole approaches the interface, the straight interface is
curved due to the forward atomic flow in front of the vortex dipole
($|\psi_1|^2 / n_0$ at $\tilde t = 0$ in Fig.~\ref{f:penet2}).
The vortex dipole then significantly deforms the interface
($|\psi_j|^2 / n_0$ at $\tilde t = 140$ and $\tilde t = 160$ in
Fig.~\ref{f:penet2}), which creates a new vortex dipole in component 2.
The vortex dipole created in component 2 then propagates in component 2,
whose cores are occupied by component 1 ($\tilde t = 300$ in
Fig.~\ref{f:penet2}).
Generation of such a ``coreless vortex dipole'' by a moving potential is
reported in Ref.~\cite{Gautam}.
If the incident velocity of the vortex dipole is slightly slower, a
greater fraction of component 1 is taken away from the interface, which
forms an elliptic ''bubble'' of component 1 moving through component 2, as
in Fig.~1 of Ref.~\cite{Sasaki}.

\begin{figure}[t]
\includegraphics[width=8cm]{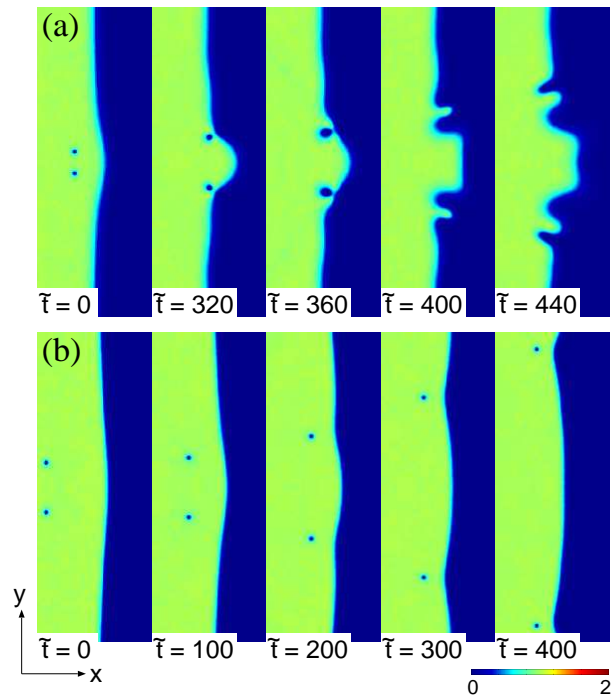}
\caption{
(Color online) Dynamics of normalized density profile $|\psi_1|^2 / n_0$
of component 1 for (a) $\Delta = 0.05$ and $v_{\rm in} = 0.088 v_{\rm s}$
and (b) $\Delta = 0.2$ and $v_{\rm in} = 0.039 v_{\rm s}$.
The field of view of each panel is $60 \xi \times 160 \xi$.
}
\label{f:waveref}
\end{figure}
Figure~\ref{f:waveref} (a) shows the dynamics for an incident velocity
$v_{\rm in}$ slower than that in Fig.~\ref{f:penet2} with the same
$\Delta$.
As the vortex dipole approaches the interface, the distance between the
vortices increases and the interface is upheaved [$\tilde t =
320$-$360$ in Fig.~\ref{f:waveref} (a)].
When the vortices touch the interface, the interface is disturbed and the
disturbance propagates along the interface [$\tilde t = 400$--$440$ in
Fig.~\ref{f:waveref} (a)].
In these dynamics, vortices are not transferred to component 2, and the
vortex penetration as in Figs.~\ref{f:penet1} and \ref{f:penet2} does not
occur.

In Fig.~\ref{f:waveref} (b), the velocity of the vortex dipole is slower
and the width of the interface is thinner than for Fig.~\ref{f:waveref}
(a).
Near the interface, the vortex and antivortex separate and move along the 
interface in opposite directions.
This behavior of vortices is similar to that of classical point vortices
near a rigid wall.
For an inviscid, incompressible, and irrotatinal fluid, the trajectories
of point vortices are given by~\cite{Lamb}
\begin{equation} \label{lamb}
\frac{1}{x^2} + \frac{1}{y^2} = \frac{4}{d^2},
\end{equation}
where $d$ is the distance between incident vortices and the wall is
located at $x = 0$ or $y = 0$.
According to Eq.~(\ref{lamb}), the distance between the trajectories and
the wall approaches $d / 2$ asymptotically, which roughly agrees with
the trajectories of the vortices in Fig.~\ref{f:waveref} (b).

\begin{figure}[t]
\includegraphics[width=8.5cm]{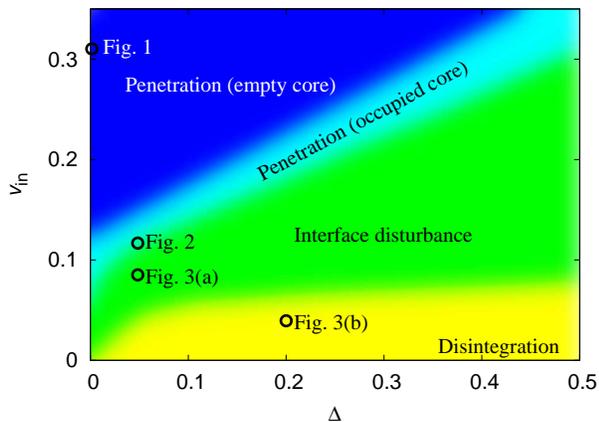}
\caption{
(Color online) Dependence of the dynamics of a vortex dipole on $\Delta$
and $v_{\rm in}$.
The region of ``Penetration (empty core)'' corresponds to the penetration
dynamics of a vortex dipole as shown in Fig.~\ref{f:penet1}.
In the region of ``Penetration (occupied core)'' the cores of a vortex
dipole after the penetration are occupied by component 1, as in
Fig.~\ref{f:penet2}.
The regions of ``Interface disturbance'' and ``Disintegration'' correspond
to the dynamics shown in Figs.~\ref{f:waveref} (a) and \ref{f:waveref}
(b), respectively.
The circles indicate the parameters used in
Figs.~\ref{f:penet1}-\ref{f:waveref}.
}
\label{f:diagram}
\end{figure}
A vortex dipole moving toward an interface thus shows a variety of dynamics
as shown in Figs.~\ref{f:penet1}-\ref{f:waveref}, which depend on the
incident velocity $v_{\rm in}$ of a vortex dipole and the parameter
$\Delta$.
Figure~\ref{f:diagram} shows the parameter dependence of the dynamics of a
vortex dipole.
The penetration of a vortex dipole across the interface occurs for large
$v_{\rm in}$ and small $\Delta$.
For small $v_{\rm in}$ and large $\Delta$, a vortex dipole behaves as if
the interface is a rigid boundary.
The region in which a vortex dipole disappears, disturbing the
interface, is located between these regions.

The parameter dependence in Fig.~\ref{f:diagram} can be understood
qualitatively.
A vortex dipole cannot penetrate the interface, when the energy of the
vortex dipole $E_{\rm vd} \simeq 2\pi \hbar^2 n_0 m^{-1} \log (d /
\xi)$~\cite{Donnely} is much smaller than the energy to deform the
interface $E_{\rm interface} \sim \sigma d$, where $\sigma$ is the
interfacial tension coefficient.
Using the expression for the interfacial tension coefficient in
Refs.~\cite{Barankov,Schae}, $\sigma = g n_0^2 \xi \Delta^{1/2}$, which is
valid for $\Delta \lesssim 0.3$, the inequality $E_{\rm vd} \ll E_{\rm
interface}$ reduces to
\begin{equation} \label{ineq}
w \ll d,
\end{equation}
where $w \sim \xi / \sqrt{\Delta}$ is the width of the interface and $\xi
\lesssim d$ is assumed.
Equation~(\ref{ineq}) indicates that the penetration of a vortex dipole is
prohibited for large $\Delta$ and small $v_{\rm vd}$, which is in
agreement with Fig.~\ref{f:diagram}.
We also expect that the penetration of a vortex dipole is prohibited when
the velocity $v_{\rm cw}$ of the capillary wave on the interface is much
faster than the velocity of a vortex dipole $v_{\rm vd} \simeq \hbar / (m
d)$, since the disturbance of the interface spreads out rapidly before the
vortex dipole is transferred across the interface, as shown in
Fig.~\ref{f:waveref} (a).
The dispersion relation of the capillary wave, $\omega^2 = \sigma k^3 / (2
m n_0)$, gives $v_{\rm cw} \sim [\sigma k / (m n_0)]^{1/2}$, and the
inequality $v_{\rm cw} \gg v_{\rm vd}$ again leads to Eq.~(\ref{ineq}).
In fluid mechanics, the ratio of inertial force to the surface or
interfacial tension force is called the Weber number, defined by ${\rm We}
= \rho v^2 \ell / \sigma$, where $\rho$ is mass density, $v$ is
characteristic velocity, and $\ell$ is characteristic length.
The substitution of $\rho = m n_0$, $v = v_{\rm vd}$, and $\ell = d$ gives
${\rm We} \sim w / d$, and therefore the penetration of a vortex dipole is
prohibited for ${\rm We} \ll 1$.

\begin{figure}[t]
\includegraphics[width=8cm]{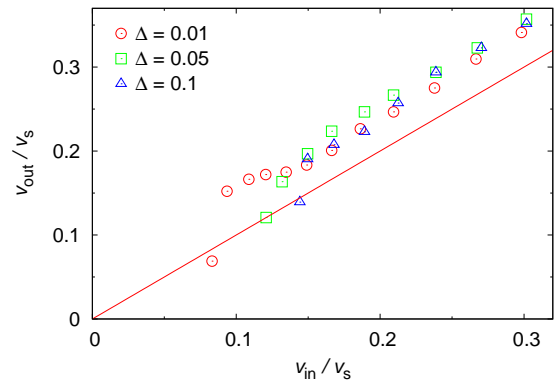}
\caption{
(Color online) Incident velocity $v_{\rm in}$ of a vortex dipole versus
outgoing velocity $v_{\rm out}$ after the penetration across the
interface for $\Delta = 0.01$ (circles), $0.05$ (squares), and $0.1$
(triangles).
The line indicates $v_{\rm in} = v_{\rm out}$.
}
\label{f:vrel}
\end{figure}
Figure~\ref{f:vrel} shows the relation between the incident velocity
$v_{\rm in}$ and outgoing velocity $v_{\rm out}$ of a vortex dipole before
and after the penetration of the interface.
For a sufficiently large incident velocity, the outgoing velocity exceeds
the incident velocity.
This is understood from the relation between the energy $E_{\rm vd}$ and
velocity $v_{\rm vd}$ of a vortex dipole,
\begin{equation}
E_{\rm vd} \sim \frac{2 \pi n_0 \hbar^2}{m} \log \frac{d}{\xi}
\simeq \frac{2 \pi n_0 \hbar^2}{m} \log \frac{\hbar}{m \xi v_{\rm vd}},
\end{equation}
i.e., the velocity $v_{\rm vd}$ is larger for smaller energy $E_{\rm
vd}$.
When passing through the interface, a vortex dipole loses energy by
disturbing the interface, and therefore the velocity increases.
In Fig.~\ref{f:vrel}, $v_{\rm out}$ rapidly falls to $\lesssim v_{\rm in}$
for a small $v_{\rm in}$, which corresponds to the ``penetration (occupied
core)'' region in Fig.~\ref{f:diagram}.
In this parameter region, the energy is given by
\begin{equation} \label{evd}
E_{\rm vd} \sim \frac{2 \pi n_0 \hbar^2}{m} \log \frac{\hbar}{m \xi v_{\rm
out}} + \frac{1}{2} m_{\rm core} v_{\rm out}^2,
\end{equation}
where $m_{\rm core}$ is the mass of component 1 occupying the cores of a
vortex dipole after the penetration.
Since $m_{\rm core}$ rapidly increases with a decrease in $v_{\rm in}$ in
this parameter region, and since the $v_{\rm out}$ dependence of the
right-hand side of Eq.~(\ref{evd}) is dominated by the second term,
$v_{\rm out}$ rapidly falls with a decrease in $v_{\rm in}$.

\begin{figure}[t]
\includegraphics[width=8cm]{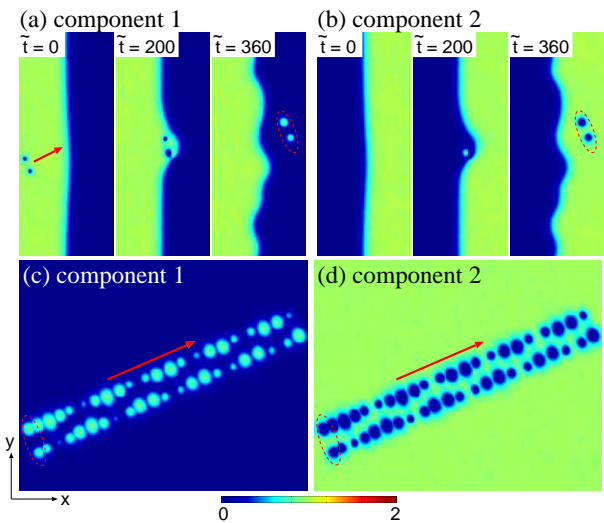}
\caption{
(Color online) (a), (b) Snapshots and (c), (d) stroboscopic images of the
normalized density profiles $|\psi_j|^2$ for $\Delta = 0.05$, $v = 0.13
v_{\rm s}$, and the incident angle, $\tan^{-1} 0.4$.
In (c) and (d), the images from $\tilde t = 360$ to $\tilde t = 1200$ are
superimposed at time intervals of $\Delta \tilde t = 40$.
The images in the dashed circles in (a) and (b) correspond to those in (c)
and (d), respectively.
The direction of vortex dipole propagation is indicated by the arrows.
The field of view of each panel is $50 \xi \times 120 \xi$ in (a) and (b),
and $120 \xi \times 80 \xi$ in (c) and (d).
}
\label{f:blink}
\end{figure}
We also examined various incident angles and found an interesting
phenomenon.
Figure~\ref{f:blink} shows the dynamics for oblique incidence of a vortex
dipole.
As shown in Figs.~\ref{f:blink} (a) and \ref{f:blink} (b), the vortex
dipole penetrates the interface in a manner similar to
Fig.~\ref{f:penet2}.
After that, a fraction of component 1 occupying the vortex cores of
component 2 oscillates between the two cores [Fig.~\ref{f:blink} (c)].
This phenomenon is due to the tunneling of a fraction of component 1 in an
effective double well potential produced by the vortex cores of component
2.
If the distance between a vortex and antivortex of a vortex dipole after
the penetration is larger, the tunneling rate is smaller and no
oscillation is observed in the relevant time scale even when only one core
is occupied by component 1 (data not shown).

\section{Dynamics of vortex dipoles in a trapped system}
\label{s:trap}

We next study the dynamics of a two-component BEC confined in a quasi-2D
axisymmetric harmonic trap $m_j \omega^2 (x^2 + y^2) / 2$.
The system is tightly confined in the $z$ direction by a harmonic
potential $m_j \omega_z^2 z^2 / 2$, and the effective interaction
coefficient in Eq.~(\ref{geff}) has the form $g_{jj'} = [\omega_z /
(2 \pi \omega)]^{1/2} G_{jj'}$.
We assume that the trap frequencies are the same for both components and
that the gravitational sag is negligible.
In the following simulations, we use $\omega = 2 \pi \times 25$ Hz and
$\omega_z = 2 \pi \times 1.25$ kHz.
We employ the $|F = 1, m_F = -1 \rangle$ state of $^{87}{\rm Rb}$ for
component 1 and $|F = 2, m_F = -2 \rangle$ state of $^{85}{\rm Rb}$ for
component 2, where $a_{11} = 99 a_{\rm B}$ and $a_{12} = 213 a_{\rm B}$
with $a_{\rm B}$ being the Bohr radius.
In the experiment reported in Ref.~\cite{Papp}, the scattering length of
$^{85}{\rm Rb}$ atoms, $a_{22}$, was controlled using the magnetic
Feshbach resonance.
We assume here that $a_{22}$ is tuned to $a_{22} = 250 a_{\rm B} >
a_{11}$, and component 2 surrounds component 1 in the ground state.
The total number of atoms is $4 \times 10^4$ with an equal population in
each component.

\begin{figure}[t]
\includegraphics[width=8.0cm]{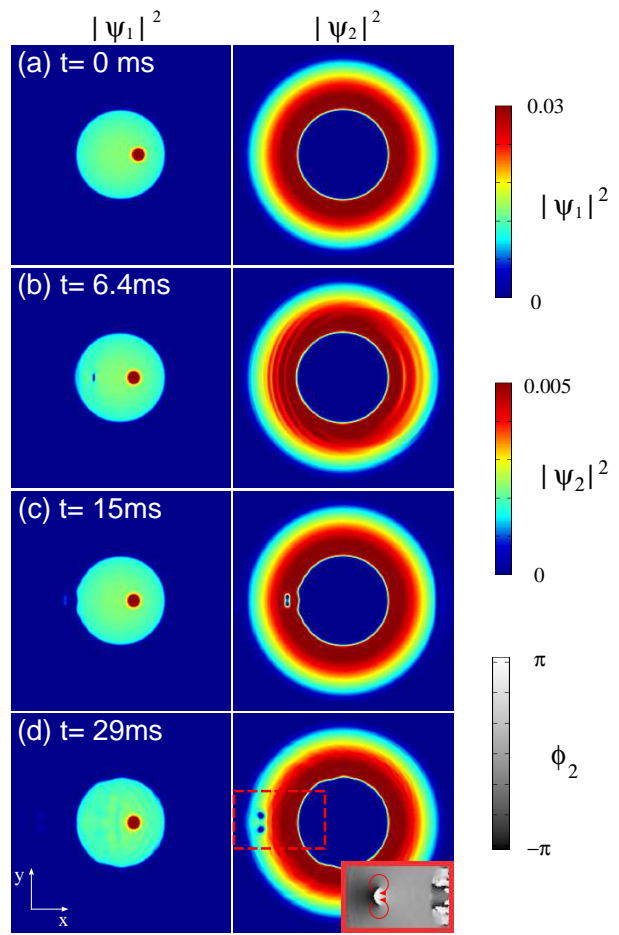}
\caption{
(Color online) Snapshots of the density profiles $|\psi_j|^2$ of the
dynamics for $T = 1.0$ ms.
The inset magnifies the phase profile ${\rm arg} \, \psi_2$ in the dashed
square.
The arrows in the inset indicate the directions of circulation.
The unit of density is $8.4 \times 10^{11}$ ${\rm cm}^{-2}$.
The field of view of each panel is $54.3 \times 54.3$ $\mu {\rm m}$.
}
\label{f:trap1}
\end{figure}
We create a vortex dipole by the same method~\cite{Aioi} as in the
previous section, i.e., the attractive Gaussian potential in
Eq.~(\ref{Gaussian}) is applied to the inner component (component 1).
The Gaussian potential with intensity $V_0 = -200 \hbar \omega$ and
waist $w = 1.09$ $\mu {\rm m}$ is moved as
\begin{equation}
x_0(t) = \left\{ \begin{array}{cc} X_0 - u t & (0 \leq t \leq T) \\
X_0 - u T & (t > T), \end{array} \right.
\end{equation}
with $X_0 = 4.36$ $\mu {\rm m}$, $u = 1.1$ ${\rm mm} / {\rm s}$, and $y_0
= 0$.
The velocity of a vortex dipole is controlled by varying $T$.
We first prepare the ground state for $x_0 = X_0$ by the imaginary-time
propagation method [Fig.~\ref{f:trap1} (a)], and then switch to real-time
propagation.

Figure~\ref{f:trap1} shows the time evolution of the system for $T = 1.0$
ms, for which a vortex dipole with velocity $\simeq 1.2$ ${\rm mm} / {\rm
s}$ is created and launched in the $-x$ direction [Fig.~\ref{f:trap1}
(b)].
The vortex dipole then penetrates the interface [Fig.~\ref{f:trap1} (c)].
After the penetration, the cores of the vortex dipole are slightly
occupied by component 1 [see the left panel of Fig.~\ref{f:trap1} (c)].
When the vortex dipole reaches the edge of the condensate
[Fig.~\ref{f:trap1} (d)], the vortices disintegrate and move in
the opposite directions along the circular edge (data not shown), as
observed in the experiment in Ref.~\cite{Neely}.

\begin{figure}[t]
\includegraphics[width=8cm]{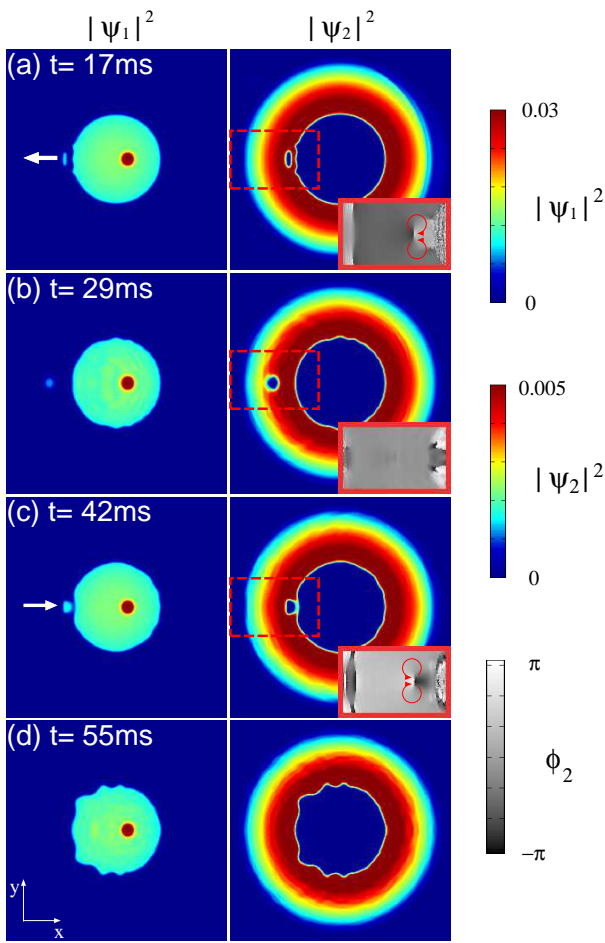}
\caption{
(Color online) Snapshots of the density profiles $|\psi_j|^2$ of the
dynamics for $T = 1.4$ ms.
The white arrows indicate the directions of propagation.
The insets magnify the phase profile ${\rm arg} \, \psi_2$ in the dashed
squares.
The arrows in the insets indicate the directions of circulation.
The unit of density is $8.4 \times 10^{11}$ ${\rm cm}^{-2}$.
The field of view of each panel is $54.3 \times 54.3$ $\mu {\rm m}$.
}
\label{f:trap2}
\end{figure}
Figure~\ref{f:trap2} shows the case of $T = 1.4$ ms, which produces a vortex
dipole with velocity $\simeq 1.1$ ${\rm mm} / {\rm s}$.
The vortex dipole penetrates the interface [Fig.~\ref{f:trap2} (a)], and
the pair of vortices merges, in which component 1 contained in the cores
also merges to become a droplet [Fig.~\ref{f:trap2} (b)].
The droplet then turns back to the center, forming a vortex--antivortex
pair in component 2 [Fig.~\ref{f:trap2} (c)].
The directions of circulation of the vortex pair in Fig.~\ref{f:trap2} (c)
are opposite to those in Fig.~\ref{f:trap2} (a).
When the droplet reaches the inner component, it eventually disappears and
the interfacial wave remains [Fig.~\ref{f:trap2} (d)].

The behaviors of the vortices in Fig.~\ref{f:trap2} can be understood as
follows.
The fraction of component 1 dragged into component 2 experiences a force
towards the center.
When the vortex dipole moves outwards [Fig.~\ref{f:trap2} (a)], it
experiences a force in the direction opposite to the propagation, and the
pair of vortices approach each other due to the Magnus effect (see
Fig.~9 of Ref.~\cite{Aioi}).
When the droplet is pushed [Fig.~\ref{f:trap2} (c)], the flow around the
droplet forms a vortex dipole.
If the droplet propagated further without reaching the interface, it
would split into two droplets (as in Fig.~2 of Ref.~\cite{Sasaki}).

\begin{figure}[t]
\includegraphics[width=8cm]{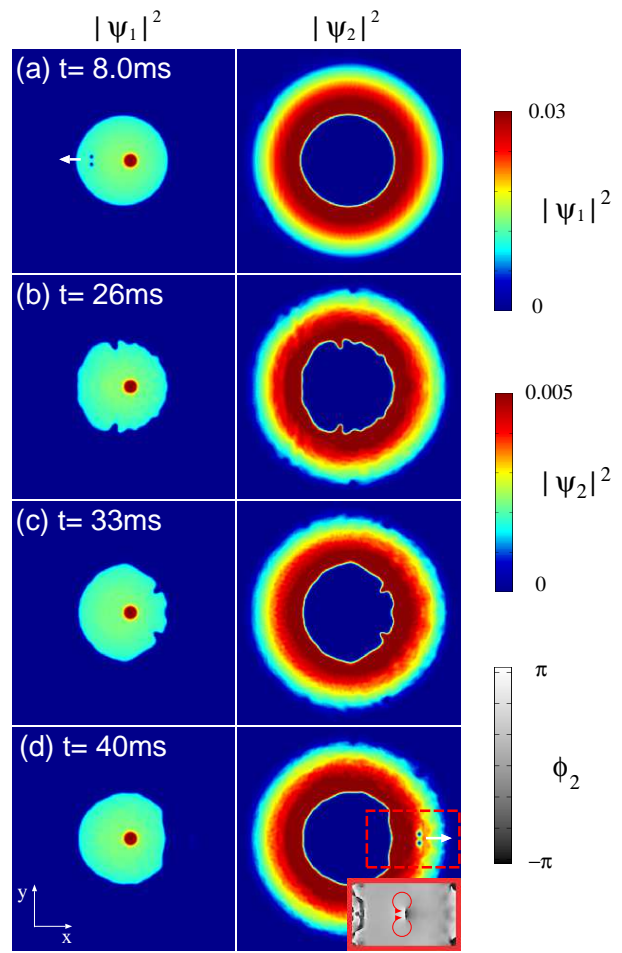}
\caption{
(Color online) Snapshots of the density profiles $|\psi_j|^2$ of the
dynamics for $T = 2.4$ ms.
The white arrow indicates the direction of propagation.
The inset magnifies the phase profile ${\rm arg} \, \psi_2$ in the dashed
square.
The arrows in the insets indicate the directions of circulation.
The unit of density is $8.4 \times 10^{11}$ ${\rm cm}^{-2}$.
The field of view of each panel is $54.3 \times 54.3$ $\mu {\rm m}$.
}
\label{f:trap3}
\end{figure}
Figure~\ref{f:trap3} shows more complicated dynamics, where a vortex
dipole with velocity $\simeq 0.9$ ${\rm mm} / {\rm s}$ is produced with
$T = 2.4$ ms.
In this case, the penetration does not occur and the vortex dipole
disappears at the interface.
The interface is disturbed and the disturbance propagates along the
interface [Figs.~\ref{f:trap3} (b) and \ref{f:trap3} (c)].
Interestingly, the disturbance refocuses at the opposite side of the
circular interface and a vortex dipole is created in component 2, which
propagates in the $+x$ direction [Fig.~\ref{f:trap3} (d)].

\section{Conclusions}
\label{s:conc}

We have investigated the dynamics of quantized vortex dipoles in
phase-separated two-component BECs.
Solving the GP equation numerically, we found that a vortex dipole can
penetrate an interface between the two components (say, from component 1
to component 2), in which quantized vortices in component 1 are
transferred to component 2 (Figs.~\ref{f:penet1} and \ref{f:penet2}).
The cores of the transmitted vortex dipole in component 2 are almost empty
(Fig.~\ref{f:penet1}) or occupied by component 1 (Fig.~\ref{f:penet2}).
When the incident velocity is slow or the width of the interface is thin,
a vortex dipole cannot penetrate the interface and the vortex dipole
disappears, disturbing the interface [Fig.~\ref{f:waveref} (a)], or the
vortex and antivortex disintegrate and move along the interface
[Fig.~\ref{f:waveref} (b)].
Through systematic numerical simulations, we obtained the parameter
dependence of the dynamics (Fig.~\ref{f:diagram}) and the relation between
incident and outgoing velocities (Fig.~\ref{f:vrel}).
When a vortex dipole penetrates the interface at an oblique angle, the
cores of the vortex dipole in component 2 are occupied by component 1
asymmetrically, followed by oscillation of component 1 between the cores
due to tunneling.

We also found a variety of dynamics for a trapped system
(Figs.~\ref{f:trap1}-\ref{f:trap3}).
After the penetration of the interface, vortices of a vortex dipole
disintegrate and move around (Fig.~\ref{f:trap1}), or change to a bubble,
falling back to the inner component (Fig.~\ref{f:trap2}).
When a vortex dipole cannot penetrate the interface, it disturbs the
interface and disappears.
In some cases, the disturbance of the circular interface focuses at the
opposite side and reproduces a vortex dipole (Fig.~\ref{f:trap3}).
These predicted phenomena can be observed in, e.g., a Feshbach-controlled
$^{85}{\rm Rb}$-$^{87}{\rm Rb}$ BEC confined in a tight pancake-shaped
trap.

\begin{acknowledgments}
This work was supported by Grants-in-Aid for Scientific
Research (No.\ 22340116 and No.\ 23540464) from the Ministry of Education,
Culture, Sports, Science and Technology of Japan.
\end{acknowledgments}

\end{document}